\documentclass[aps,prb,twocolumn,superscriptaddress,showpacs]{revtex4}
\usepackage{graphicx}
\usepackage{amsmath}
\usepackage{epsfig}

\newcommand{\beq}{\begin{eqnarray}}
\newcommand{\eeq}{\end{eqnarray}}

\begin{document}

\title{Entanglement swapping and testing quantum steering into the past via
collective decay}
\author{Yueh-Nan Chen}
\email{yuehnan@mail.ncku.edu.tw}
\affiliation{Department of Physics and National Center for Theoretical Sciences, National
Cheng-Kung University, Tainan 701, Taiwan}
\author{Shin-Liang Chen}
\affiliation{Department of Physics and National Center for Theoretical Sciences, National
Cheng-Kung University, Tainan 701, Taiwan}
\author{Neill Lambert}
\affiliation{CEMS, RIKEN, Saitama, 351-0198, Japan}
\author{Che-Ming Li}
\affiliation{Department of Engineering Science, National Cheng-Kung University, Tainan
City 701, Taiwan}
\author{Guang-Yin Chen}
\affiliation{Department of Physics, National Chung Hsing University, Taichung 402, Taiwan}
\author{Franco Nori}
\affiliation{CEMS, RIKEN, Saitama, 351-0198, Japan}
\affiliation{Physics Department, University of Michigan, Ann Arbor, MI 48109-1040, USA}
\affiliation{Department of Physics, Korea University, Seoul 136-713, Korea}
\date{\today }

\begin{abstract}
We propose a scheme to realize entanglement swapping via superradiance,
entangling two distant cavities without a direct interaction. The successful
Bell-state-measurement outcomes are performed naturally by the
electromagnetic reservoir, and we show how, using a quantum trajectory
method, the non-local properties of the state obtained after the swapping
procedure can be verified by the steering inequality. Furthermore, we
discuss how the unsuccessful measurement outcomes can be used in an
experiment of delayed-choice entanglement swapping. An extension of testing
the quantum steering inequality with the observers at three different times
is also considered.
\end{abstract}

\pacs{03.67.Bg, 42.50.Nn, 03.65.Ud}
\maketitle

\section{INTRODUCTION}

Entanglement swapping \cite{Ekert} is a procedure to create entanglement
between two qubits which have never directly interacted with each other, and
has been demonstrated experimentally by Pan\textit{\ et al.} \cite{Pan}.
Together with quantum memories \cite{Duan}, one can, in principle, use
entanglement swapping to build quantum repeaters \cite{Briegel} to overcome
the decoherence problem in quantum communication \cite{Gisin}. In fact,
entanglement swapping can also be viewed as a special example of quantum
teleportation \cite{Bennett} if the unknown state is replaced by an
entangled state~\cite{tel}. Very recently, the notion of delayed-choice
entanglement swapping was experimentally demonstrated by Ma \textit{et al}. %
\cite{Ma}, an idea first considered in Peres' \cite{Peres} and Cohen's \cite%
{Cohen} gedanken experiments. The intriguing feature of this mechanism is
that the entanglement is made `a posteriori', after the entangled qubits
have been measured.

%To create entanglement between two qubits a direct interaction between
%them is usually employed.
Another route to create entanglement between two qubits is to make use of
the common environment %through which the two qubits are entangled %
\cite{reservoir}, i.e. via which they can effectively interact. Taking this
one step further, Chen \textit{et al}. \cite{YNChen} combined the concept of
teleportation and a common environment and proposed a method to teleport
charge qubits via a common reservoir using the super-radiance effect. The
necessary Bell-state measurements are performed naturally by the collective
decay, i.e. the sub- and superradiance channels \cite{Dicke}. An analysis of
the fidelity and the success probability was further examined \cite{Clemens}
using the quantum trajectory method.

In this work, we propose a scheme to accomplish entanglement swapping via
collective decay, effectively swapping the entanglement between two atoms
into entanglement between two distant cavities. The steering inequality \cite%
{steering} is utilized to verify the non-local properties of the final state
obtained between the two distant cavities. We further point out that the
unsuccessful outcomes, naturally occurring from the superradiance effect we
employ here, may be useful in a delayed-choice entanglement-swapping
experiment. Finally, we consider the scenario of quantum steering into the
past when the observers perform their measurements at three different times.
\begin{figure}[tbp]
\includegraphics[width=9cm]{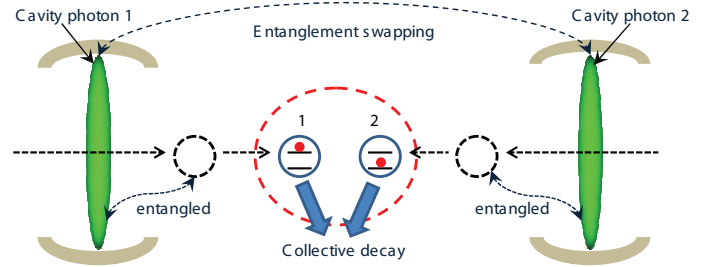}
\caption{(Color online) Schematic description of entanglement swapping via
superradiance. First, the entanglement is generated between the $j$-th atom
and its corresponding cavity. The next step is to trap the two atoms, so
that they decay collectively. If the measurement outcome is a single photon,
the entanglement swapping can be achieved.}
\end{figure}

\section{ENTANGLEMENT\ SWAPPING\ VIA\ SUPERRADIANCE}

It is well known that highly-entangled states can be `naturally' generated
via collective spontaneous decay \cite{Dicke}. For two identical qubits
interacting with a common photon reservoir, separated by a distance shorter
than the emitted radiation wavelength, entanglement appears in the two
intermediate states, $\left| S_{0}\right\rangle =(\left| +-\right\rangle
-\left| -+\right\rangle )/\sqrt{2}$ and $\left| T_{0}\right\rangle =(\left|
+-\right\rangle +\left| -+\right\rangle )/\sqrt{2}$, of the two decay
channels \cite{Devoe} from the excited state $\left| T_{1}\right\rangle
=\left| ++\right\rangle $ to the ground state $\left| T_{-1}\right\rangle
=\left| --\right\rangle $. Armed with this, we now propose how to accomplish
the entanglement swapping protocol through such collective decay processes.

We first consider two identical two-level atoms each passing through a
separate cavity, as shown in Fig. 1. At this stage the atom-cavity system 1
and atom-cavity system 2 are well separated from each other. In the strong
atom-cavity coupling regime the interactions between the atoms and their
respective cavities can be written as \cite{CQED}
\begin{equation}
H^{\prime }=\hbar g_{j}(\sigma _{j,+}b_{j}^{-}+\sigma _{j,-}b_{j}^{+}),
\end{equation}%
where $j=1$ or $2$, $g_{j}$ is the atom-cavity coupling strength, $%
b_{j}^{\pm }$ and the Pauli matrices $\sigma _{j,\pm }$ are the cavity
photon and the atom operators respectively. With the appropriate preparation
of the initial states and control of the passage times through the cavities,
the singlet entangled states
\begin{equation}
(\left| 0\right\rangle _{c,j}\left| +\right\rangle _{j}-\left|
1\right\rangle _{c,j}\left| -\right\rangle _{j})/\sqrt{2},
\end{equation}%
are prepared between the $j$-th atom and its corresponding cavity. Here, $%
\left| 0\right\rangle _{c,j}$ ($\left| 1\right\rangle _{c,j}$) and $\left|
+\right\rangle _{j}$ ($\left| -\right\rangle _{j}$) represent the $j$-th
cavity with no (one) photon and the $j$-th atom in its excited (ground)
state, respectively.

Our goal is to use these entangled atom-cavity states and the phenomena of
collective decay, as well as post-selection, to perform entanglement
swapping and entangle the photonic states of the two distant cavities. To
achieve this, the next step is to remove the atoms from the cavities and
trap the two atoms close together so that they experience superradiant
collective-decay processes in the common electromagnetic environment. This
is a collective decay phenomenon which is enhanced when the inter-atom
distance is much shorter than the wavelength of the emitted photon. One can
also have a similar enhanced effect by placing the two atoms at the
anti-nodes of a cavity \cite{reservoir}. Before the collective decay occurs,
the total wavefunction of the combined atom-cavity systems can be written as
\begin{eqnarray}
\left| \Psi \right\rangle &=&\left[ \frac{1}{2}(\left| 0\right\rangle
_{c,1}\left| +\right\rangle _{1}-\left| 1\right\rangle _{c,1}\left|
-\right\rangle _{1})\otimes (\left| 0\right\rangle _{c,2}\left|
+\right\rangle _{2}\right.  \notag \\
&&\left. -\left| 1\right\rangle _{c,2}\left| -\right\rangle _{2})\right]
\notag \\
&=&\frac{1}{2}\left[ \left| 0\right\rangle _{c,1}\left| 0\right\rangle
_{c,2}\otimes \left| T_{1}\right\rangle _{12}+\left| 1\right\rangle
_{c,1}\left| 1\right\rangle _{c,2}\otimes \left| T_{-1}\right\rangle
_{12}\right.  \notag \\
&&\left. +\left( \left| 0\right\rangle _{c,2}\left| 1\right\rangle
_{c,1}-\left| 1\right\rangle _{c,2}\left| 0\right\rangle _{c,1}\right)
\otimes \left| S_{0}\right\rangle _{12}\right.  \notag \\
&&\left. -\left( \left| 0\right\rangle _{c,2}\left| 1\right\rangle
_{c,1}+\left| 1\right\rangle _{c,2}\left| 0\right\rangle _{c,1}\right)
\otimes \left| T_{0}\right\rangle _{12}\right] .
\end{eqnarray}%
Assuming that all photonic decay processes from the two atoms can be
observed (inefficient detection is introduced below), there are four
possible outcomes due to the collective decay: zero photons emitted ($\left|
T_{-1}\right\rangle _{12}$), two photons emitted ($\left| T_{1}\right\rangle
_{12}$), and finally just one photon via sub-radiant channel ($\left|
S_{0}\right\rangle _{12}$), or one photon via super-radiant channel ($\left|
T_{0}\right\rangle _{12}$). If the measurement outcome is a single photon,
i.e. via either $\left| S_{0}\right\rangle _{12}$ or $\left|
T_{0}\right\rangle _{12}$, then entanglement swapping can be achieved,
provided that the sub- and super-radiant decay can be distinguished. As
pointed out in Ref. [\onlinecite{YNChen}], the momentum of the emitted
photon $\overrightarrow{k}$\ depends on the separation of the two atoms $%
\overrightarrow{r}$, i.e. $\overrightarrow{k}\cdot $\ $\overrightarrow{r}$\ $%
=0$\ or $\pi $\ corresponds to the emission of a super- or sub-radiant
photon, respectively. Therefore, to distinguish between the sub- and
super-radiant photons one can place the detectors at the appropriate angles.

One can use the recently proposed steering inequality \cite{steering} or
Bell-CHSH inequality \cite{Horodecki} to verify the non-local properties of
the state obtained after the swapping procedure. The density operator of the
state obtained after the post-selection measurement can be easily calculated
by using the quantum trajectory method \cite{Clemens}, where it can be
described as a probabilistic mixture of different measurement outcomes:%
\begin{equation}
\rho (t)=\sum_{i}p_{i}\left| \psi _{i}(t)\right\rangle \left\langle \psi
_{i}(t)\right| ,
\end{equation}%
where $i$ denotes the events, or measurements, of photodetection and $\left|
\psi _{i}(t)\right\rangle $ is the pure state conditioned on this event. For
example, the total master equation of the evolution of two two-atoms, and
their collective decay phenomena, without post-selection, can be written as%
\begin{equation}
\frac{d}{dt}\rho =-\frac{i}{\hbar }[H,\rho ]+\sum_{i}^{n}(2\widehat{J}%
_{i}\rho \widehat{J}_{i}^{\dagger }-\widehat{J}_{i}^{\dagger }\widehat{J}%
_{i}\rho -\rho \widehat{J}_{i}^{\dagger }\widehat{J}_{i}).
\end{equation}%
$H$ is any Hermitian Hamiltonian evolution (e.g., dipole-dipole interactions
between the atoms, or external magnetic fields). Here we neglect such terms
and set $H=0$. The photon-emission-event operators are
\begin{eqnarray}
\widehat{J}_{1}^{{}} &=&\sqrt{\frac{\gamma +\Gamma }{2}}(\sigma
_{1,-}+\sigma _{2,-}), \\
\widehat{J}_{2}^{{}} &=&\sqrt{\frac{\gamma -\Gamma }{2}}(\sigma
_{1,-}-\sigma _{2,-}).
\end{eqnarray}%
We assume the two atoms are separated by a distance $d$ and that the
wavelength of the emitted light is $\lambda $. Thus, $\gamma $ is the
spontaneous emission rate for a single atom and $\Gamma =\frac{\sin (2\pi
d/\lambda )}{(2\pi d/\lambda )}\gamma $.

To obtain the post-selected state, one can unravel the evolution (see Ref.~[%
\onlinecite{Clemens}] for a full description of a similar derivation for a
teleportation scheme driven by superradiance), and then the un-normalized
state of the two-atoms is given by:%
\begin{eqnarray}
\overline{\left| \psi _{i}(t)\right\rangle } &=&e^{-iH_{B}(t-t_{n})}\widehat{%
J}_{i_{n}}e^{-iH_{B}(t_{n}-t_{n-1})}\cdot \cdot \cdot \widehat{J}%
_{i_{2}}\times  \notag \\
&&e^{-iH_{B}(t_{2}-t_{1})}\widehat{J}_{i_{1}}e^{-iH_{B}t_{1}}\left| \psi
(0)\right\rangle ,
\end{eqnarray}%
with the effective non-Hermitian Hamiltonian written as
\begin{equation}
H_{B}=-i\hbar (\widehat{J}_{1}^{\dagger }\widehat{J}_{1}^{{}}+\widehat{J}%
_{2}^{\dagger }\widehat{J}_{2}^{{}}).
\end{equation}

As pointed out in Ref.~[\onlinecite{Clemens}], the sub- and super-radiant
outcomes can also be distinguished in a statistical sense. In this case, one
notes that the state $\left| S_{0}\right\rangle _{12}$\ ($\left|
T_{0}\right\rangle _{12}$) favors the longer (shorter) emission times.
Following the derivation in Ref.~[\onlinecite{Clemens}], one can define a
crossover time $t_{1}^{\ast }$, such that the swapped state should be
corrected (i.e., $\left| 0\right\rangle _{c,2}\left| 1\right\rangle
_{c,1}+\left| 1\right\rangle _{c,2}\left| 0\right\rangle _{c,1}\rightarrow
\left| 0\right\rangle _{c,2}\left| 1\right\rangle _{c,1}-\left|
1\right\rangle _{c,2}\left| 0\right\rangle _{c,1}$) for a single photon
emission at time $t_{1}<t_{1}^{\ast }$. For $t_{1}>t_{1}^{\ast }$, there is
no need to correct the swapped state.\emph{\ }We now assume that after the
atoms are brought together one waits a time $T$ and retains the state when
exactly one photon, emitted by the atoms, was detected in the period $T$
(with detection efficiency $\eta $). After averaging over all such detection
events (distinguished by the statistical scheme mentioned above), and
tracing out the atomic states \cite{Clemens}, the remaining two-cavity state
is%
\begin{eqnarray}
\rho _{c}(\eta ,T) &=&\eta \rho _{c}(T)+\frac{\eta (1-\eta )}{4}\left|
0\right\rangle _{c,1}\left| 0\right\rangle _{c,2}\otimes \text{ }%
_{c,2}\left\langle 0\right| _{c,1}\left\langle 0\right|  \notag \\
&&\times \Bigg\{\frac{2(\gamma ^{2}+\Gamma ^{2})}{\gamma ^{2}-\Gamma ^{2}}%
(1+e^{-2\gamma T})-\frac{2}{\kappa }e^{-(\gamma +\Gamma )T}  \notag \\
&&-2\kappa e^{-(\gamma -\Gamma )T}-\frac{4\Gamma ^{2}}{\gamma ^{2}-\Gamma
^{2}}(1-e^{-2\gamma T})\Bigg\},
\end{eqnarray}%
where%
\begin{eqnarray}
\rho _{c}(T) &=&\left| 0\right\rangle _{c,1}\left| 0\right\rangle
_{c,2}\otimes \text{ }_{c,2}\left\langle 0\right| _{c,1}\left\langle
0\right| \Big\{\frac{1}{4\kappa }[e^{-(\gamma +\Gamma )T}  \notag \\
&&-e^{-2\gamma T}]+\frac{\kappa }{4}[e^{-(\gamma -\Gamma )T}-e^{-2\gamma T}]%
\Big\}  \notag \\
&&+\frac{1}{4}\left| \psi ^{+}\right\rangle \left\langle \psi ^{+}\right| %
\big\{1-\kappa ^{(\gamma +\Gamma )/2\Gamma }  \notag \\
&&+\kappa ^{(\gamma -\Gamma )/2\Gamma }-e^{-(\gamma -\Gamma )T}\big\}  \notag
\\
&&+\frac{1}{4}\left| \psi ^{-}\right\rangle \left\langle \psi ^{-}\right| %
\big\{1-\kappa ^{(\gamma -\Gamma )/2\Gamma }  \notag \\
&&+\kappa ^{(\gamma +\Gamma )/2\Gamma }-e^{-(\gamma +\Gamma )T}\big\}.
\end{eqnarray}%
Here,
\begin{eqnarray}
&&\kappa \equiv (\gamma -\Gamma )/(\gamma +\Gamma ), \\
&&\left| \psi ^{-}\right\rangle =(\left| 0\right\rangle _{c,2}\left|
1\right\rangle _{c,1}-\left| 1\right\rangle _{c,2}\left| 0\right\rangle
_{c,1})/\sqrt{2}, \\
&&\left| \psi ^{+}\right\rangle =(\left| 0\right\rangle _{c,2}\left|
1\right\rangle _{c,1}+\left| 1\right\rangle _{c,2}\left| 0\right\rangle
_{c,1})/\sqrt{2}.
\end{eqnarray}

To test the non-local properties of the resultant cavity state $\rho
_{c}(\eta ,T)$, we use both the steering inequality \cite{steering} and the
maximum value of a Bell inequality violation \cite{Horodecki}. For the
steering inequality, the correlated measurements observed by Alice and Bob
on the two cavities are described by the probability distribution $%
P(B_{i}=b, $ $A_{i}=a)$ with $b=\pm 1,$or $0$, and $a=\pm 1$. If the two
cavities are not entangled, the steering inequality is written as \cite%
{steering}%
\begin{equation}
S_{N}\equiv \sum_{i=1}^{N}E\left[ \left\langle \widehat{A}_{i}\right\rangle
_{B_{i}}^{2}\right] \leq 1,
\end{equation}%
where $N$($=2$ or $3$) is the number of mutually-unbiased measurements that
Alice implements on her qubit (cavity), and
\begin{equation}
E\left[ \left\langle \widehat{A}_{i}\right\rangle _{B_{i}}^{2}\right] \equiv
\sum_{b=\pm 1,0}P(B_{i}=b)\left\langle \widehat{A}_{i}\right\rangle
_{B_{i}=b}^{2}
\end{equation}%
with Alice's expectation value for a measured (conditioned on Bob's result)
defined as
\begin{equation}
\left\langle \widehat{A}_{i}\right\rangle _{B_{i}=b}\equiv P(A_{i}=+1\mid
B_{i}=b)-P(A_{i}=-1\mid B_{i}=b).
\end{equation}%
For the maximum value of a Bell inequality violation, the Bell operator
associated with the CHSH inequality has the following form \cite{Horodecki},%
\begin{equation}
\widehat{B}_{CHSH}\equiv \widehat{a}\cdot \widehat{\sigma }\otimes (\widehat{%
b}+\widehat{b}^{\prime })\cdot \widehat{\sigma }+\widehat{a}^{\prime }\cdot
\widehat{\sigma }\otimes (\widehat{b}-\widehat{b}^{\prime })\cdot \widehat{%
\sigma },
\end{equation}%
where $\widehat{a}$, $\widehat{a}^{\prime }$, $\widehat{b}$, $\widehat{b}%
^{\prime }$ are unit vectors in \textbf{R}$^{3}$. Here, $\widehat{a}\cdot
\widehat{\sigma }\equiv \sum_{i=1}^{3}a_{i}\sigma _{i},$ where $\sigma _{i}$
are the standard Pauli matrices. If the two cavities are not entangled, the
CHSH inequality of the two-cavity state $\rho $ obeys%
\begin{equation}
\left| \left\langle \widehat{B}_{CHSH}\right\rangle _{\rho }\right| =\left|
\text{Tr}(\rho \widehat{B}_{CHSH})\right| \leq 2\text{.}
\end{equation}%
The maximum value of the CHSH inequality is given by
\begin{equation}
B_{\max }=\max_{\widehat{a},\widehat{a}^{\prime },\widehat{b},\widehat{b}%
^{\prime }}\text{Tr}(\rho \widehat{B}_{CHSH})\text{.}
\end{equation}
Thus, violations of the steering parameter $S_{N}$ or the maximum value of
the CHSH inequality $B_{\max }$ can mean that the resultant cavity state is
entangled.

In Fig. 2(a), we plot the steering parameter $S_{N}$ and the maximum value
of the CHSH inequality $B_{\max }$ for various photon detection
efficiencies. When decreasing the photon detection efficiency $\eta $ below
0.79, the steering inequality (dashed curves) is still violated while the
Bell-CHSH one (solid curves) is not. For the inter-atomic distance $d$, the
range for the violation of the steering parameter is larger than that for
the mean value of the Bell observable. In Fig. 2(b), we also plot $S_{N}$
and $B_{\max }$ as a function of the waiting time $T$. As seen, if the
waiting time $T$ is too short, it is possible that the inequalities are not
violated. This coincides with the results obtained in Ref.~[%
\onlinecite{Clemens}]: The longer the waiting time $T$, the higher the
success probability. Besides, the results in Fig. 2(a) and (b) all show that
the steering inequality has better tolerance in examining the non-local
properties of the entangled states \cite{steering}.
\begin{figure}[tbp]
\includegraphics[width=8.5cm]{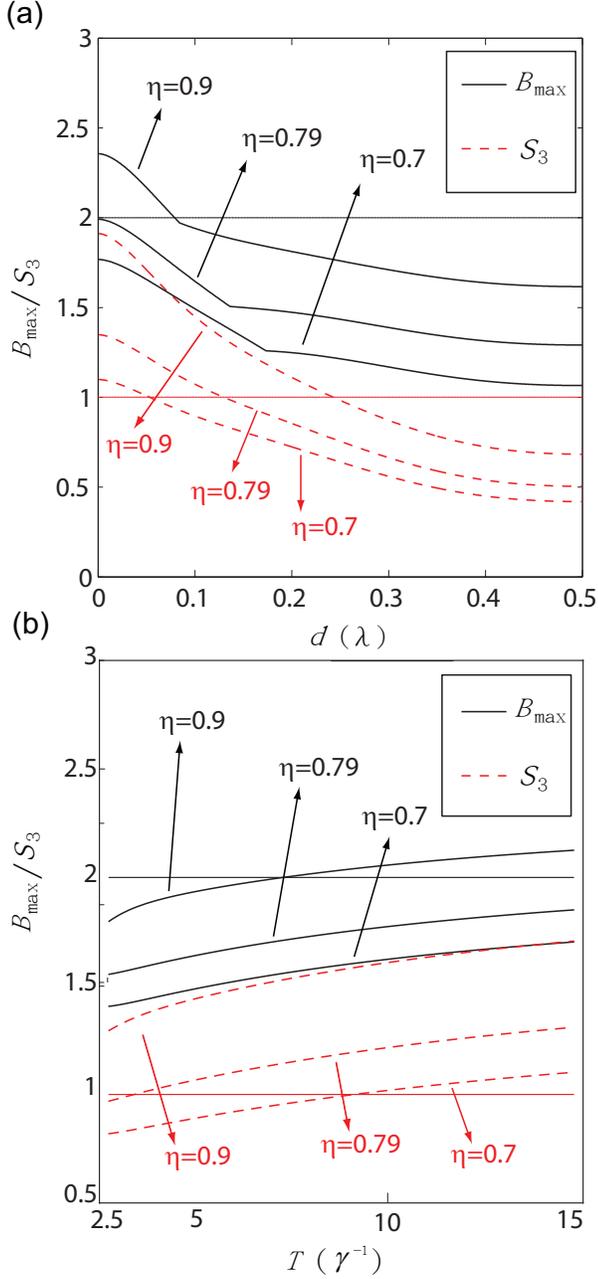} %
\caption{(Color online) Testing the steering and Bell-CHSH inequalities for
the two-cavity state $\protect\rho _{c}(T)$ after entanglement swapping. The
solid curves and dashed curves represent the results of the maximum value of
Bell inequality $B_{\max }$ and the steering parameter $S_{3}$,
respectively. The horizontal black line is the Bell-CHSH inequality bound
and the horizontal red line is the $S_{3}$ bound. In plotting the figure
(a), we set the waiting time $T$ $=(5/\protect\gamma )$, and the
inter-atomic distance $d$ is in units of the wavelength $\protect\lambda $
of the emitted photon. In plotting the figure (b), we set the the
inter-atomic distance $d=0.1\protect\lambda $.}
\end{figure}

%An experimental issue that would need\cite{Gisin} to be addressed to measure the steering inequality in a real experiment is how to realize the
%measurements of the cavity photon states in mutually unbiased bases. Ultimately this depends
%on the specific physical realization, for example with superconducting qubits and microwave transmission
%line cavities, one could utilize an ancilla qubit coherently coupled
%to the cavity
%\cite{You}. As demonstrated in Ref.~[\onlinecite{transfer}], the cavity
%state could then be transferred to the atomic state if the qubit is tuned to be resonant
%with the cavity, after which one could perform the mutually unbiased measurements on
%the qubit.

\section{DELAYED-CHOICE ENTANGLEMENT SWAPPING}

From Eq.~(3), naively one knows there is a 50\% chance that this protocol
may fail, i.e., one may obtain the outcomes $\left| T_{-1}\right\rangle
_{12} $ or $\left| T_{1}\right\rangle _{12}$. This drawback is actually
useful if one wishes to implement delayed-choice entanglement swapping. To
illustrate this, let us start with Peres'\cite{Peres} original gedanken
experiment. The joint state of a pair of singlets (particles $a$\ and $b$\
and particles $c$ and $d$) takes the form%
\begin{equation}
\left| \Psi \right\rangle _{abcd}=\left| \psi ^{-}\right\rangle _{ab}\otimes
\left| \psi ^{-}\right\rangle _{cd},
\end{equation}%
where $\left| \psi ^{-}\right\rangle _{ab}=(\left| \uparrow \right\rangle
_{a}\left| \downarrow \right\rangle _{b}-\left| \downarrow \right\rangle
_{a}\left| \uparrow \right\rangle _{b})/\sqrt{2}$ and likewise for $\left|
\psi ^{-}\right\rangle _{cd}$. Here, $\left| \uparrow \right\rangle _{k}$
and $\left| \downarrow \right\rangle _{k}$ are the two spin states of the
particles $k=a,$ $b,$ $c,$ $d$. Equation (21) can be rewritten in the basis
of Bell states of particles $a$ and $d$, and $b$ and $c$:

\begin{eqnarray}
|\Psi \rangle _{abcd} &=&\frac{1}{2}\left( \left| \psi ^{+}\right\rangle
_{ad}\otimes \left| \psi ^{+}\right\rangle _{bc}-\left| \psi
^{-}\right\rangle _{ad}\otimes \left| \psi ^{-}\right\rangle _{bc}\right.
\notag \\
&&\left. -\left| \phi ^{+}\right\rangle _{ad}\otimes \left| \phi
^{+}\right\rangle _{bc}+\left| \phi ^{-}\right\rangle _{ad}\otimes \left|
\phi ^{-}\right\rangle _{bc}\right) ,  \notag \\
&&
\end{eqnarray}%
where $|\psi ^{+}\rangle _{ad}=(\left| \uparrow \right\rangle _{a}\left|
\downarrow \right\rangle _{d}+\left| \downarrow \right\rangle _{a}\left|
\uparrow \right\rangle _{d})/\sqrt{2}$ and $\left| \phi ^{\pm }\right\rangle
_{ad}=(\left| \uparrow \right\rangle _{a}\left| \uparrow \right\rangle
_{d}\pm \left| \downarrow \right\rangle _{a}\left| \downarrow \right\rangle
_{d})/\sqrt{2}$ are the symmetric Bell states (and likewise for particles $b$
and $c$). In the normal scheme of delayed-choice entanglement swapping,
particles $a$ and $d$ are sent to Alice and Bob. The third observer, Eve,
performs the Bell-state measurement on particles $b$ and $c$ \textit{after}
Alice and Bob have measured the values of their spin components (randomly
chosen along arbitrary directions). With Eq. (22), Alice and Bob can then
sort their data into four subsets according to the measurement outcomes of
Eve. If Alice and Bob test the Bell's inequality with only the data in one
of the subsets, they would find the inequality is readily violated.

\begin{figure}[tbp]
\includegraphics[width=8cm]{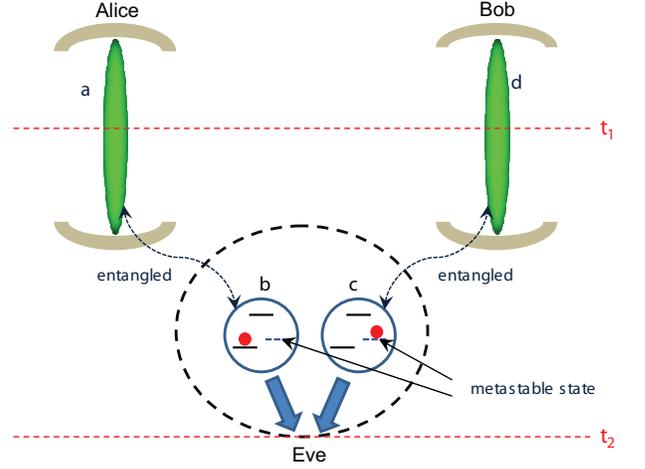}
\caption{(Color online) Delayed-choice entanglement via superradiance. When
the atoms $b$ and $c$ are trapped, laser pulses are applied to store the
information in the metastable states with a sufficiently longer
dephasing/decay time. After the measurements are performed on the cavities ($%
a$ and $d$) at time $t_{1}$, the reverse pulses are applied to the atoms $b$
and $c$ to continue the collective decay at a later time $t_{2}$.}
\end{figure}

Very recently, Ma \textit{et al.} \cite{Ma} experimentally demonstrated the
delayed-choice entanglement swapping protocol using photons. In their
experiment, instead of choosing all of the four Bell-state measurements,
they performed measurements which either project onto entangled states or
onto the separable states ($\left| \uparrow \right\rangle _{a}\left|
\uparrow \right\rangle _{d} $ or $\left| \downarrow \right\rangle _{a}\left|
\downarrow \right\rangle _{d}$). This allows them to a \textit{posteriori}
decide on whether Alice and Bob's states are entangled or separable.
%Basically, an active, random, and delayed choice, that the photons cannot
%know in advance the setting of the future measurement, is demonstrated.

Returning to our entanglement swapping scheme in Eq. (3), although there is
a 50\% chance that the swapping may fail, the unsuccessful results ($\left|
T_{-1}\right\rangle _{12}$ and $\left| T_{1}\right\rangle _{12}$) are just
those that project the particles into the separable states \cite{Ma}, and
therefore can be used in the experiment of delayed-choice entanglement
swapping. An additional advantage here, over that of a scheme based on
purely photonic degrees of freedom, is that we can make use of other
internal metastable states of the atoms to postpone the collective decay
(Fig. 3). When the atoms $b$ and $c$ are trapped, laser pulses can be
applied to store the information in metastable states with a much longer
dephasing/decay time. After Alice and Bob have performed their measurements
on their own particles (which here are now the photonic degrees of freedom
in the cavities) at time $t_{1}$, the reverse pulses are applied to the
atoms $b$ and $c$ to continue the collective decay at a later time $t_{2}$.

\begin{figure}[tbp]
\includegraphics[width=8cm]{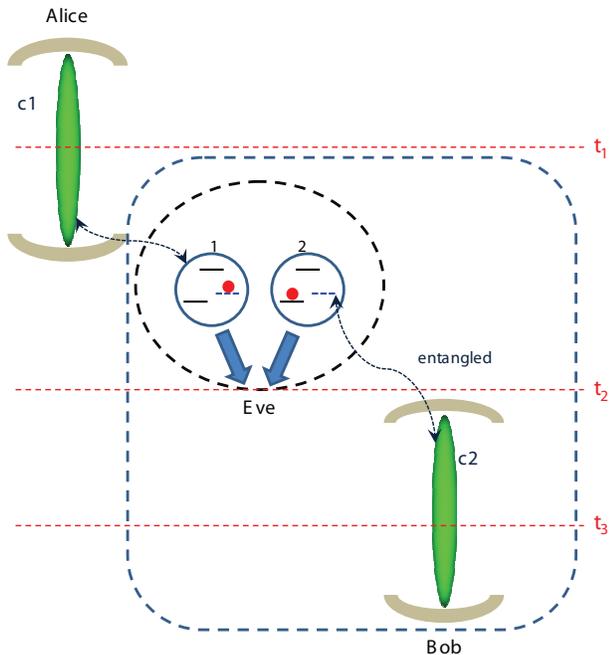}
\caption{(Color online) To test quantum steering into the past via
superradiance, the relative temporal order of the three observers should be
arranged as: $t_{1}$(Alice)$<$ $t_{2}$(Eve)$<$ $t_{3}$(Bob). If the atomic
regime and Bob's cavity are considered as a black box (blue-dashed line),
this means that Bob can steer Alice's state by his measurement.}
\end{figure}

\section{TESTING QUANTUM STEERING INTO THE PAST}

In our scheme, the Bell measurements are made `automatically' by the common
photon reservoir. Therefore, the delayed choice is intrinsically random,
i.e. in the stochastic sense of spontaneous decay. However, a loophole may
arise when Alice and Bob happen to choose their measurements both along the $%
z$-axis. If their measurement outcomes are the same ($\left| 0\right\rangle
_{c,1}\left| 0\right\rangle _{c,2}$ or $\left| 1\right\rangle _{c,1}\left|
1\right\rangle _{c,2}$), from Eq.~(3) they would know in advance that the
later joint-measurement settings can only be the separable ones, and could
never be the entangled ones. To overcome this, the relative temporal order
of the three observers should be changed: $t_{1}$(Alice) $<$ $t_{2}$(Eve) $<$
$t_{3}$(Bob) as shown in Fig. 4. In this case, even if Alice chooses her
setting along the $z$-axis, the wavefunction of Eq.~(3) may, for example,
collapse onto
\begin{eqnarray}
\left| \Psi \right\rangle _{bcd} &=&\frac{1}{2}[\left| 0\right\rangle
_{c,2}\otimes \left| T_{-1}\right\rangle _{12}  \notag \\
&&-\left| 1\right\rangle _{c,2}\otimes \left| S_{0}\right\rangle
_{12}+\left| 1\right\rangle _{c,2}\otimes \left| T_{0}\right\rangle _{12}].
\end{eqnarray}%
The later joint atomic-measurement driven by the common photon reservoir can
still either project it into the separable state ($\left|
T_{-1}\right\rangle _{12}$) or the entangled states ($\left|
S_{0}\right\rangle _{12}$, $\left| T_{0}\right\rangle _{12}$). Actually,
such an arrangement can be viewed as a special kind of quantum steering if
the atomic regime and Bob's cavity are considered as a black box \cite%
{steering}. This means Bob can steer Alice's state by his measurement, which
is conditioned on the random choice of the collective decay. Remember that
Alice's measurement is performed before Bob's measurement. Similar to the
Bell's test in Ref. [\onlinecite{Ma}], one can therefore test `quantum
steering into the past' by using the steering inequality \cite{steering}.

\begin{figure}[tbp]
\includegraphics[width=9cm]{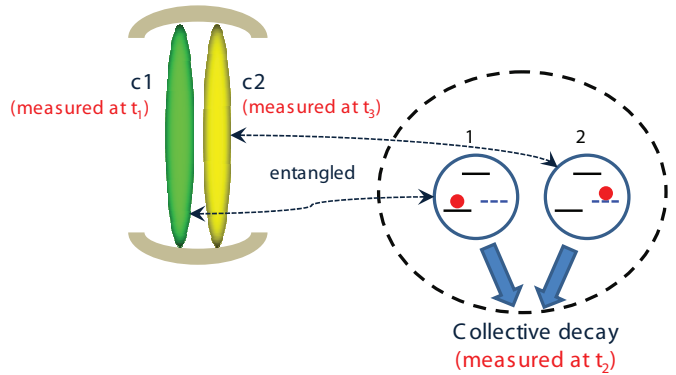}
\caption{(Color online) A simplified version of testing quantum steering
into the past. After the atom $1$ is trapped and the information is stored
in its metastable state, one then performs a measurement on cavity $c_{1}$
at time $t_{1}$. The next step is to reset the cavity and let atom $2$ pass
through it. When atom $2$ arrives at the trap, the reverse pulse is applied
to atom $1$, so that a collective decay can occur at time $t_{2}$. Then, the
measurement is performed on the cavity $c2$ at time $t_{3}.$}
\end{figure}

The above proposal may also be further simplified by using only one cavity
with two atoms. As shown in Fig. 5, one first lets the atom $1$ pass through
the cavity, and as it does so the cavity plays the role of cavity $c1$ in
the above scheme. After the atom $1$ has exited the cavity and is trapped,
and its state has been transferred to an internal metastable state, one then
performs a measurement on the photonic state in the cavity $c1$ and records
the data. The next step is to `reset' the cavity and let atom $2$ pass
through it, so that the cavity now plays the role of the cavity $c2$. When
atom $2$ arrives at the trap, the atom $1$ is transferred back to its
original state so that the collective decay can occur. Then, a measurement
is again performed on the cavity $c2$.

Following the above procedure one can obtain the violation of the steering
inequality and may have the illusion of ``self-entanglement'' since the
measurement data is produced by the same cavity. However, this is not the
case because when we reset the cavity, it (of course) becomes a new system.
The use of a single cavity simplifies a possible experimental implementation
and emphasizes the fact that the relative temporal order of the three
observers' events is irrelevant \cite{Peres, Ma}.

\section{SUMMARY}

In summary, we have proposed a scheme to accomplish entanglement swapping
via superradiance. We outlined how the swapping protocol could be combined
with collective decay to entangle two distant cavities. After post-selection
and averaging over all single-photon events, the non-local properties of the
two-cavity state were analyzed by both the steering and Bell-CHSH
inequalities. We found that the steering inequality has a better tolerance
in verifying the non-local properties of the swapping state. Furthermore, we
have also pointed out that the unsuccessful events in our scheme can be used
in a delayed-choice entanglement swapping protocol.

\section{ACKNOWLEDGEMENTS}

This work is supported partially by the National Center for Theoretical
Sciences and National Science Council, Taiwan, grant numbers NSC
101-2628-M-006-003-MY3 and NSC 100-2112-M-006-017. FN is partially supported
by the ARO, RIKEN iTHES Project, MURI Center for Dynamic Magneto-Optics,
JSPS-RFBR Contract No. 12-02-92100, Grant-in-Aid for Scientific Research
(S), MEXT Kakenhi on Quantum Cybernetics, and the JSPS via its FIRST Program.

\end{document}